\documentclass[aps,prb,twocolumn,showpacs,groupedaddress,nofootinbi]{revtex4}
\usepackage{here}
\usepackage{graphicx}

\begin{document}


\title{Magnetic phase diagram of CePt$_3$B$_{1-x}$Si$_x$}
\author{D. Rauch$^{1}$, S. S\"{u}llow$^1$, M. Bleckmann$^{1,2}$, B. Klemke$^3$, K. Kiefer$^3$, M.~S. Kim$^4$, M.~C. Aronson$^{4,5}$, E. Bauer$^6$}

\address{$^1$ Institute of Condensed Matter Physics, University of Technology Braunschweig, D-38106 Braunschweig, Germany\\
$^2$ Wehrwissenschaftliches Institut f\"ur Werk- und Betriebsstoffe WIWeB, D-85435 Erding, Germany\\
$^3$ Berlin Neutron Scattering Center, Helmholtz Zentrum Berlin, D-14109 Berlin, Germany\\
$^4$ Stony Brook University, Department of Physics and Astronomy, New York, USA\\
$^5$ Brookhaven National Laboratory, Condensed Matter Physics and Materials Science Department, New York, USA\\
$^6$ Institute of Solid State Physics, Vienna University of Technology, A-1090 Vienna, Austria}

\date{\today}

\begin{abstract}
We present a study of the main bulk properties (susceptibility, magnetization, resistivity and specific heat) of CePt$_3$B$_{1-x}$Si$_x$, an alloying system that crystallizes in a noncentrosymmetric lattice, and derive the magnetic phase diagram. The materials at the end point of the alloying series have previously been studied, with CePt$_3$B established as a material with two different magnetic phases at low temperatures (antiferromagnetic below $T_N = 7.8$\,K, weakly ferromagnetic below $T_C \approx 5$\,K), while CePt$_3$Si is a heavy fermion superconductor ($T_c = 0.75$\,K) coexisting with antiferromagnetism ($T_N = 2.2$\,K). From our experiments we conclude that the magnetic phase diagram is divided into two regions. In the region of low Si content (up to $x \sim 0.7$) the material properties resemble those of CePt$_3$B. Upon increasing the Si concentration further the magnetic ground state continuously transforms into that of CePt$_3$Si. In essence, we argue that CePt$_3$B can be understood as a low pressure variant of CePt$_3$Si.
\end{abstract}


\maketitle 

\section{Introduction}

The observation of superconductivity in the noncentrosymmetric heavy fermion compound CePt$_3$Si opened up a new field of research on superconductors,\cite{Bauer04} which nowadays stands in the focus of intense research efforts (for a review see Ref.~\cite{pfleiderer} and references therein). Most notably, subsequent to the discovery of superconductivity in CePt$_3$Si similar superconducting states were observed in closely related Ce intermetallics, {\it viz.}, CeRhSi$_3$, CeIrSi$_3$, CeCoGe$_3$ and CeIrGe$_3$.\cite{kimura,sugitani,kawai,honda} Similar to other Ce heavy fermion superconductors,\cite{steglich,jaccard,movshovich,mathur,grosche,sarrao,settai} for the noncentrosymmetric systems there is a close relation between (the suppression of) magnetic order and the appearance of superconducting states.\cite{kimura,sugitani,kawai,honda} These observations mostly were derived from studies of the ground state properties of the materials under externally applied pressure. Here, in particular, the behavior of CePt$_3$Si has been investigated quite extensively under applied pressure.\cite{yasuda,tateiwa,nicklas,aoki,nicklas10,motoyama}

At ambient pressure, CePt$_3$Si was reported to crystallize in the tetragonal noncentrosymmetric CePt$_3$B structure (space group $P4mm$),\cite{Bauer04} with lattice parameters $a = 4.072$ \AA~ and $c = 5.442$ \AA . Based on various studies by means of thermodynamic as well as microscopic techniques the system was characterized as heavy fermion ($\gamma = 0.39$ J/mole K$^2$) superconductor below $T_c = 0.75$~K (0.45 K in a high quality single crystal).\cite{Bauer04,takeuchi} Moreover, it was established that superconductivity coexists with a long-range antiferromagnetically ordered state, with an ordering wave vector ${\bf q} = (0, 0, 0.5)$ of strongly reduced moments $\mu_{ord} = 0.16 \mu_B$/Ce being detected below the N$\mathrm{\acute{e}}$el temperature $T_N = 2.2$ K.\cite{meteoki,amato} 

The superconducting state in CePt$_3$Si is believed to be of an unconventional nature, although as yet a detailed description of superconductivity has not been developed. Further, the lack of inversion symmetry causes a spin-orbit splitting of the Fermi surface, which might generate chiral spin states.\cite{hashimoto} As result of the various pressure studies, most notably it is found that the antiferromagnetic state is already suppressed at a pressure of about 0.6 GPa, while superconductivity persists up to a pressure of 1.5 GPa. Thus, for noncentrosymmetric CePt$_3$Si the appearance of superconductivity is closely linked to the suppression of magnetic order, although here quantum critical behavior has not been observed in the various physical properties.

The ternary compound CePt$_3$B is isostructural to the heavy fermion superconductor CePt$_3$Si, with lattice parameters $a = 4.003$ \AA~ and $c = 5.075$ \AA.\cite{sullow,lackner} With these - compared to CePt$_3$Si - much smaller lattice parameters it might be argued that chemical pressure effects take place here, such that in some sense CePt$_3$B would represent a high pressure variant of CePt$_3$Si. However, with the replacement of Si by B the electron count is lower by one electron in CePt$_3$B. Naively, one could argue that this should tend to weaken the hybridization strength because of a lower electron density corresponding to a reduced chemical pressure. 

Effectively, the experimentally observed physical properties of CePt$_3$B appear to lead to the conclusion that this material is a local moment magnet with much weaker electronic correlations than CePt$_3$Si. Based on thermodynamic and transport experiments it has been established that CePt$_3$B undergoes two magnetic transition at low temperatures, the first one into a presumably antiferromagnetic (AFM) state below $T_N = 7.8$ K, the latter one into a state with a weakly ferromagnetic (FM) signature below $T_C \sim 4.5 - 6$ K. Here, to account for the two magnetic phases in CePt$_3$B one would argue that there is a transition of large magnetic moments (order of magnitude $\sim \mu_B$) into an antiferromagnetic structure below $T_N$, and which transforms into a weakly ferromagnetic one below $T_C$ through canting of the magnetic moments. Within this line of thought, the canting might be a consequence of the lacking inversion symmetry, as this would give rise to an additional magnetic exchange term, the Dzyahloshinskii-Moriya interaction.\cite{dm} Here, a magnetic exchange term $\propto S_1 \times S_2$ between two spins $S_1$, $S_2$ on sites without inversion symmetry, if combined with a ferro- or antiferromagnetic coupling, produces complex magnetic states such as canted or helical structures.\cite{prokes,bak,nakanishi}

Surprisingly, in a recent study of the magnetically ordered phases of CePt$_3$B by means of neutron scattering and muon spin rotation ($\mu SR$) this scenario could not be verified.\cite{rauch} On the one hand, in $\mu SR$ experiments both transitions $T_N$ and $T_C$ have been identified as bulk transitions. As well, the muon precession frequency does suggest the presence of a fairly large ordered magnetic moment in both phases. On the other hand, in neutron powder diffraction no additional intensity from scattering in the magnetically ordered phase has been observed. As yet, this failure to detect magnetic intensity in neutron scattering is not understood.

Given that CePt$_3$Si and CePt$_3$B are isostructural, the question arises if there is a relationship between the magnetically ordered phases in both compounds. One possible route to study this topic is an alloying experiment on the series of materials CePt$_3$B$_{1-x}$Si$_x$ for $0 \leq x \leq 1$. Here, it can be studied how the antiferromagnetic state of CePt$_3$Si evolves out of that of CePt$_3$B, and if the noncentrosymmetric crystal structure plays a role in defining the magnetic ground state. In the following we will present such a study. We will characterize the structural and physical ground state properties of the alloying series CePt$_3$B$_{1-x}$Si$_x$, and in particular will discuss our data in terms of chemical pressure effects.

\section{Sample preparation and experimental techniques}

Polycrystalline samples of CePt$_3$B$_{1-x}$Si$_x$, $0 \leq x \leq 1$, have been prepared by high frequency melting the constituents in stoichiometric ratio under argon atmosphere in a water-cooled copper crucible. Subsequently the samples have been annealed at 880$^\circ$C for 14 days in evacuated quartz tubes. 

\begin{figure}[!ht]
	\centering
		\includegraphics[width=1\linewidth]{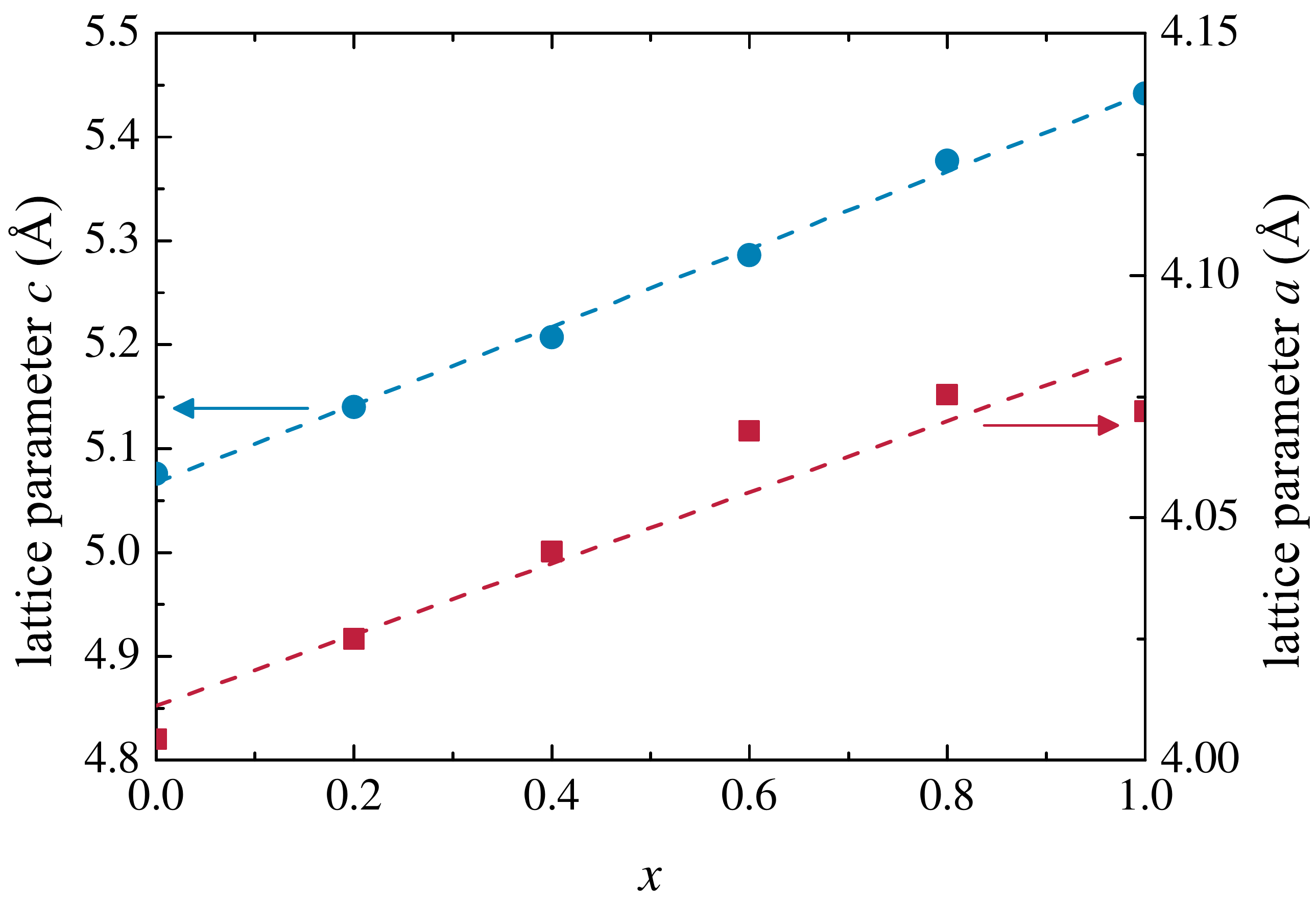}
	\caption{(Color online) Lattice parameters $a$ and $c$ versus silicon concentration for CePt$_3$B$_{1-x}$Si$_x$, 
	$0 \leq x \leq 1$.}
	\label{fig:latticepara}
\end{figure}

Metallurgically, the materials have been characterized by means of powder x-ray diffraction. All samples crystallize in the tetragonal lattice with the space group {\it P4mm}, in agreement with the Refs.~\cite{Bauer04,sullow,Sologub02}. Within experimental resolution ($\sim 10$~vol.\%) no secondary phases have been detected in the diffraction spectra. The lattice parameters $a$ and $c$ as determined from x-ray diffraction are summarized in table~\ref{tab:Temp1}. Further, Fig.~\ref{fig:latticepara} visualizes the significant increase of the lattice parameters upon replacing boron by silicon, with a basically linear evolution of the lattice parameters in accordance with Vegard's law. In the figure and the table the data from the references \cite{Bauer04,lackner} are included. Overall, replacing boron by silicon leads to an increase of the unit cell volume of about 10\%. Using the bulk modulus of 162~GPa,\cite{ohashi} this increase of the unit cell volume would correspond to a negative chemical pressure of about 16~GPa. Notably, the change of the $c$ axis parameter is much larger (about 7 \% from CePt$_3$B to CePt$_3$Si) than the $a$ axis parameter (less than 2 \%), which would indicate some anisotropy of this chemically exerted pressure.

\begin{table}[!ht]
	\begin{tabular}{l c c c c c c }\hline  
		 $x$  & ~ 0.0~\cite{lackner} & 0.2 & 0.4 & 0.6 & 0.8 & 1.0~\cite{Bauer04}\\
		\hline
		\hline 
		 $a$ (\AA) & ~ 4.004 & ~ 4.025 & ~ 4.043 & ~ 4.068 & ~ 4.075 & ~ 4.072\\
		 $c$ (\AA) & ~ 5.075 & ~ 5.140 & ~ 5.207 & ~ 5.286 & ~ 5.377 & ~ 5.442\\
		 $V$ (\AA$^3$) & ~ 81.36 & ~ 83.27 & ~ 85.11 & ~ 87.48 & ~ 89.29 & ~ 90.23\\
				\hline
		\end{tabular}
	\caption{Lattice parameters $a$, $c$ and unit cell volume $V$ of CePt$_3$B$_{1-x}$Si$_x$ for $0 \leq x \leq~1$ (space group {\it P4mm}).}
	\label{tab:Temp1}
\end{table}

The physical bulk properties of CePt$_3$B$_{1-x}$Si$_x$ have been examined by means of susceptibility, magnetization, resistivity and specific heat measurements. The susceptibility and magnetization were measured employing a commercial SQUID magnetometer, at temperatures ranging from 1.8 to 300~K in fields up to 5~T. Resistivity measurements were carried out using a standard ac four-point technique at temperatures from 1.8 to 300~K. Furthermore, specific heat measurements have been performed in a commercial calorimeter from 0.3 to 300~K at Helmholtz Zentrum Berlin (Germany) and at Brookhaven National Laboratory (USA).


\section{Experimental results}

\begin{figure}[!ht]
	\centering
		\includegraphics[width=1\linewidth]{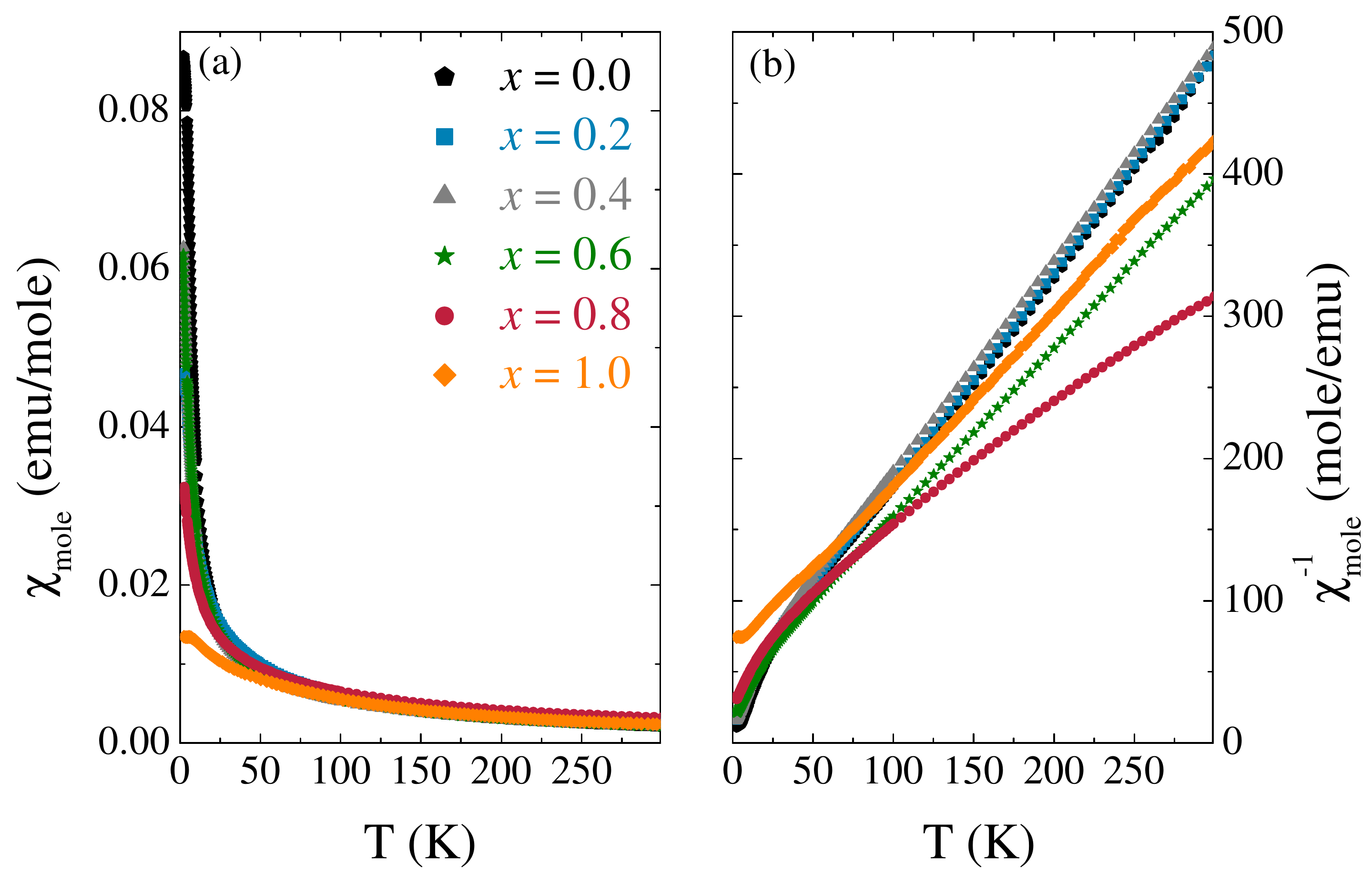}
	\caption{(Color online) Temperature dependence of (a) the susceptibility $\chi (T)$ and (b) inverse susceptibility $\chi^{-1} (T)$ of CePt$_3$B$_{1-x}$Si$_x$ for $0 \leq x \leq~1$ in a magnetic field of $B=1$~T. The data for CePt$_3$Si are taken from Ref.~\cite{Bauer05}.}
	\label{fig:suszep1}
\end{figure}

In Fig.~\ref{fig:suszep1} the temperature dependence of the magnetic susceptibility $\chi (T)$ and inverse susceptibility $\chi^{-1} (T)$ in a field $B=1$~T are depicted. At high temperatures a paramagnetic Curie-Weiss behavior is observed. The effective Ce moments $\mu_{eff}$ are derived from fits of the data between 50 and 300~K using the common expression~\cite{footnote1} $\chi = \left[ C/(T-\Theta_{CW}) \right] + \chi_0$, $C \propto \mu_{eff}^2$, and summarized in Tab.~\ref{tab:Temp2}. These data indicate a stable Ce$^{3+}$ state at high temperatures for all compositions CePt$_3$B$_{1-x}$Si$_x$. Further, the Curie-Weiss temperatures $\Theta_{CW}$ indicate predominant antiferromagnetic interactions, which increase with silicon concentration $x$. Moreover, also the Kondo temperature $T_K$ depends on the magnetic coupling strength $J \propto \Theta_{CW}$, which thus increases as well.

\begin{table}[!ht]
	\centering
	  \begin{tabular}{l c c c c c c c }\hline  
		 $x$  & & ~ 0.0 & ~ 0.2 & ~ 0.4 & ~ 0.6 & ~ 0.8 & ~ 1.0\\
		\hline
		\hline   
		$T_N$~(K) & $\chi$ & ~ 8.1 & ~ 5.6 & ~ 4.3 & ~ 2.9 & ~ 2.2 & ~ 2.2 \\
		$T_N$~(K) & $\rho$ & ~ 7.8 & ~ 6.1 & ~ 4.5 & ~ 2.8 & ~ 2.6 & ~ 2.2 \\
		$T_N$~(K) & $C_p$ & ~ 7.8 & ~ 5.3 & ~ 4.3 & ~ 2.9 & ~ 2.6 & ~ 2.2 \\ \hline
		$T_C$~(K) & $\chi$ & ~ 5.6 & ~ 2.4 & ~ 2.0 & ~ - & ~ - & ~ - \\
		$T_{C}$~(K) & $C_p$ & ~ $\approx$ 4.5 & ~ 1.8 & ~ 1.7 & ~ 1.6 & ~ - & ~ - \\ \hline
		$\Theta_{CW}$~(K) & $\chi$ & ~ -26 & ~ -28 & ~ -29 & ~ -31 & ~ -44 & ~ -46 \\
		$\mu_{eff}$~($\mu_{B}$) & $\chi$ & ~ 2.39 & ~ 2.39 & ~ 2.34 & ~ 2.56 & ~ 2.58 & ~ 2.54\\
		\hline
		$T_{CEF}$~(K) & $\rho$ & ~ 150 & ~ 143 & ~ 140 & ~ 130 & ~ 131 & ~ 119 \\
		$T_{K}$~(K) & $\rho$ & $\approx$~7 & ~ 12 & ~ 19 & ~ 18 & ~ 16 & ~ 11 \\ \hline
		$\gamma$ & $C_p$ & ~ 57 & ~ 151 & ~ 295 & ~ 337 & ~ 349 & ~ 390 \\
		(mJ/mole K$^2$) & & & & & & & \\
		$S_{mag}(T=T_N)$ & $C_p$ & ~ 0.78 & ~ 0.74 & ~ 0.57 & ~ 0.52 & ~ 0.48 & ~ 0.22 \\
		($R \cdot \ln{2}$) & & & & & & & \\ \hline
		\end{tabular}
\caption{Magnetic transition temperatures and characteristic physical parameters of CePt$_3$B$_{1-x}$Si$_x$, as determined from susceptibility $\chi$, resistivity $\rho$ and specific heat $C_p$: antiferromagnetic transition temperature $T_N$, ferromagnetic transition temperature $T_C$, Curie-Weiss temperature $\Theta_{CW}$, effective magnetic moment $\mu_{eff}$, characteristic temperature of the crystal field splitting $T_{CEF}$, Kondo temperature $T_{K}$, Sommerfeld coefficient $\gamma$ and magnetic entropy $S_{mag}$ at $T = T_N$. Values for CePt$_3$B and CePt$_3$Si are taken from the Refs.~\cite{sullow,lackner,Bauer04}.}
	\label{tab:Temp2}
\end{table}

\begin{figure}[!ht]
	\centering
		\includegraphics[width=1\linewidth]{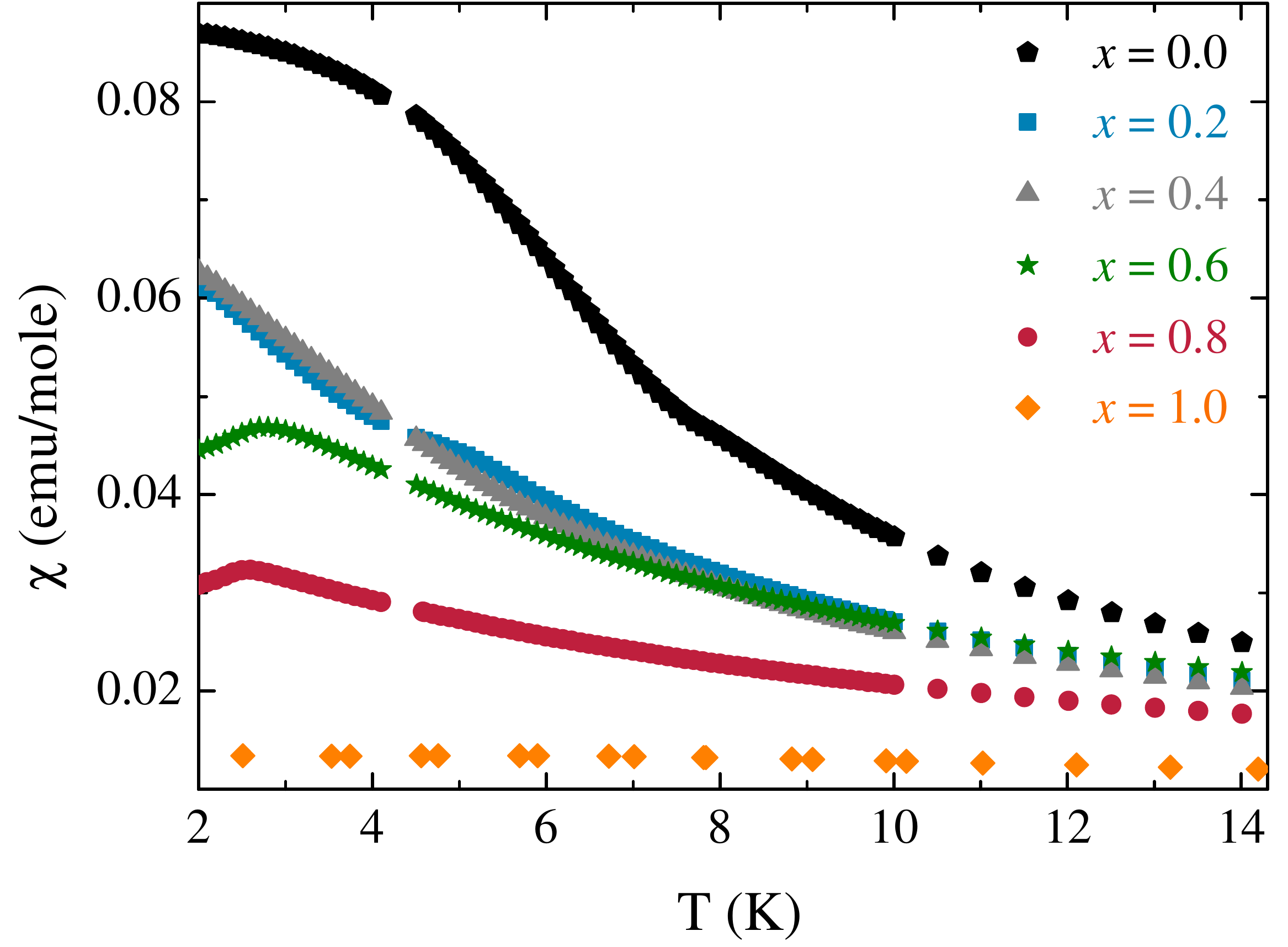}
	\caption{(Color online) Low temperature regime of the susceptibility $\chi (T)$ of CePt$_3$B$_{1-x}$Si$_x$, $0 \leq x \leq~1$, in a magnetic field of $B=1$~T. The data for CePt$_3$Si are taken from Ref.~\cite{Bauer05}.}
	\label{fig:suszep}
\end{figure}

As demonstrated in Fig.~\ref{fig:suszep}, at low temperatures deviations from Curie-Weiss behavior become apparent, denoting transitions into long-range ordered states. Transition temperatures are determined as anomalies in plots $\chi T$ vs. $T$, with $T_C$ identified as maximum of $\chi T$, and $T_N$ as inflection point. This way, the transition temperatures for CePt$_3$B are determined as $T_N = 8.1$~K and $T_C = 5.6$~K, in good agreement with the Refs. \cite{lackner,sullow}. Next, for $x$ up to $0.4$ two magnetic phase transitions, an antiferromagnetic and a ferromagnetic one, are identified. In contrast, for a larger silicon amount $x$ the ferromagnetic transition seems to have disappeared, while the antiferromagnetic transition persists for all $x$, with $T_N$ decreasing to $2.2$~K in CePt$_3$Si. The values $T_N$ and $T_C$ as determined from the susceptibility are also summarized in Tab.~\ref{tab:Temp2}.

\begin{figure}[!ht]
	\centering
		\includegraphics[width=1\linewidth]{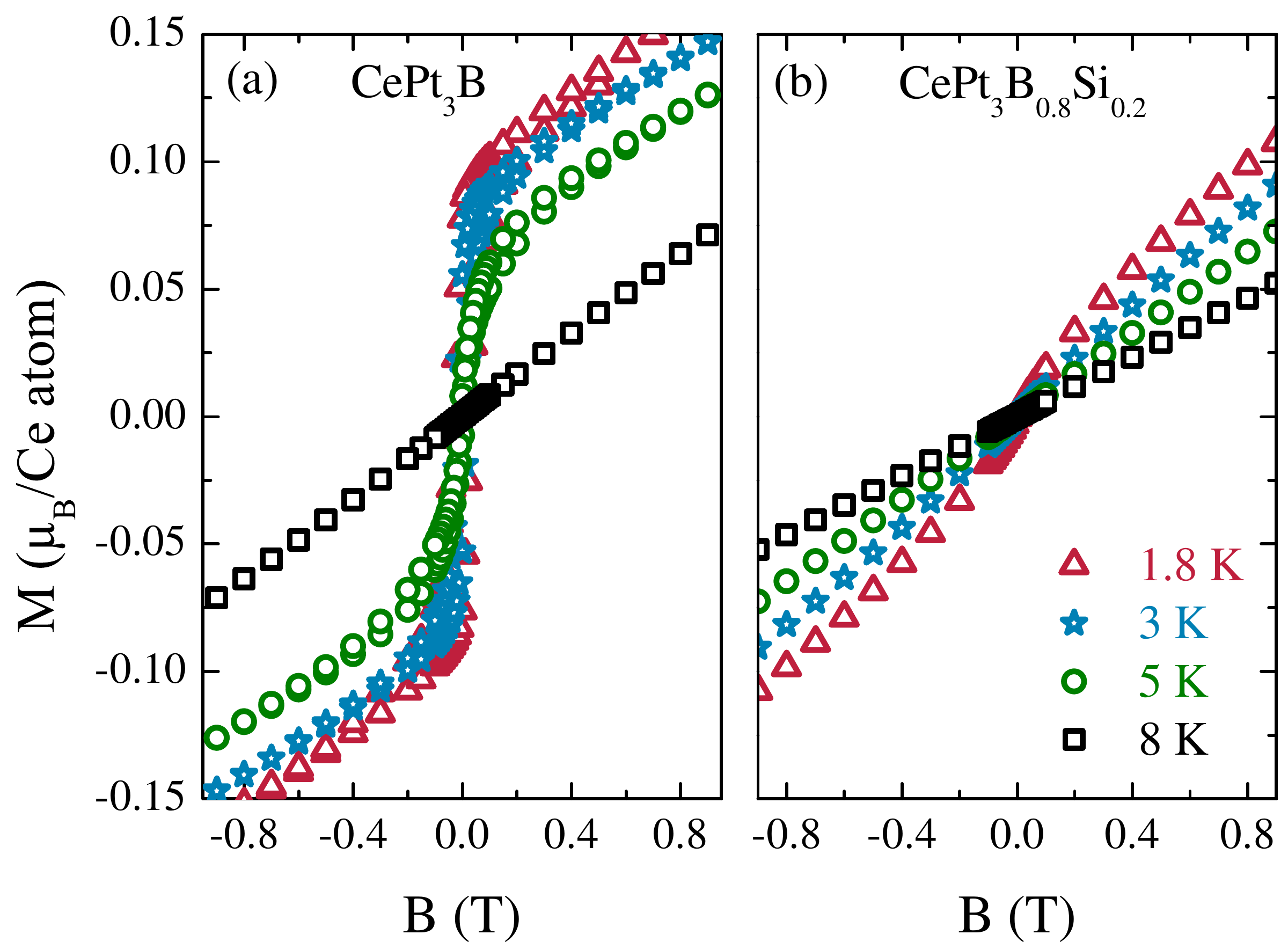}
	\caption{(Color online) Magnetic field dependence of the magnetization of CePt$_3$B (a) and CePt$_3$B$_{0.8}$Si$_{0.2}$ (b).}
	\label{fig:magnet}
\end{figure}

Furthermore, magnetization measurements at low temperatures have been carried out on CePt$_3$B$_{1-x}$Si$_x$ for $0 \leq x \leq 0.8$. CePt$_3$B exhibits weak ferromagnetic hysteresis below $T_C$ (extrapolated remanent ferromagnetic moment for $T \rightarrow 0$~K: $0.09 \mu_B$/Ce atom), in good agreement with Refs. \cite{sullow, rauch} (see Fig.~\ref{fig:magnet}(a)). Conversely, for the samples CePt$_3$B$_{1-x}$Si$_x$, $x \neq 0$, no ferromagnetic hysteresis is observed (cf. Fig.~\ref{fig:magnet}(b)). However, the samples $x = 0.2$ and 0.4 both display a weakly ferromagnetic shape of the magnetization curve at $1.8$~K (estimated remanent ferromagnetic moment at $1.8$~K: $\sim 0.02 \mu_B$/Ce atom), consistent with the observation of a second phase transition in the $\chi$ measurements.
As an example, the corresponding data are shown for CePt$_3$B$_{0.8}$Si$_{0.2}$ in Fig.~\ref{fig:magnet}(b).

\begin{figure}[!ht]
	\centering
		\includegraphics[width=1\linewidth]{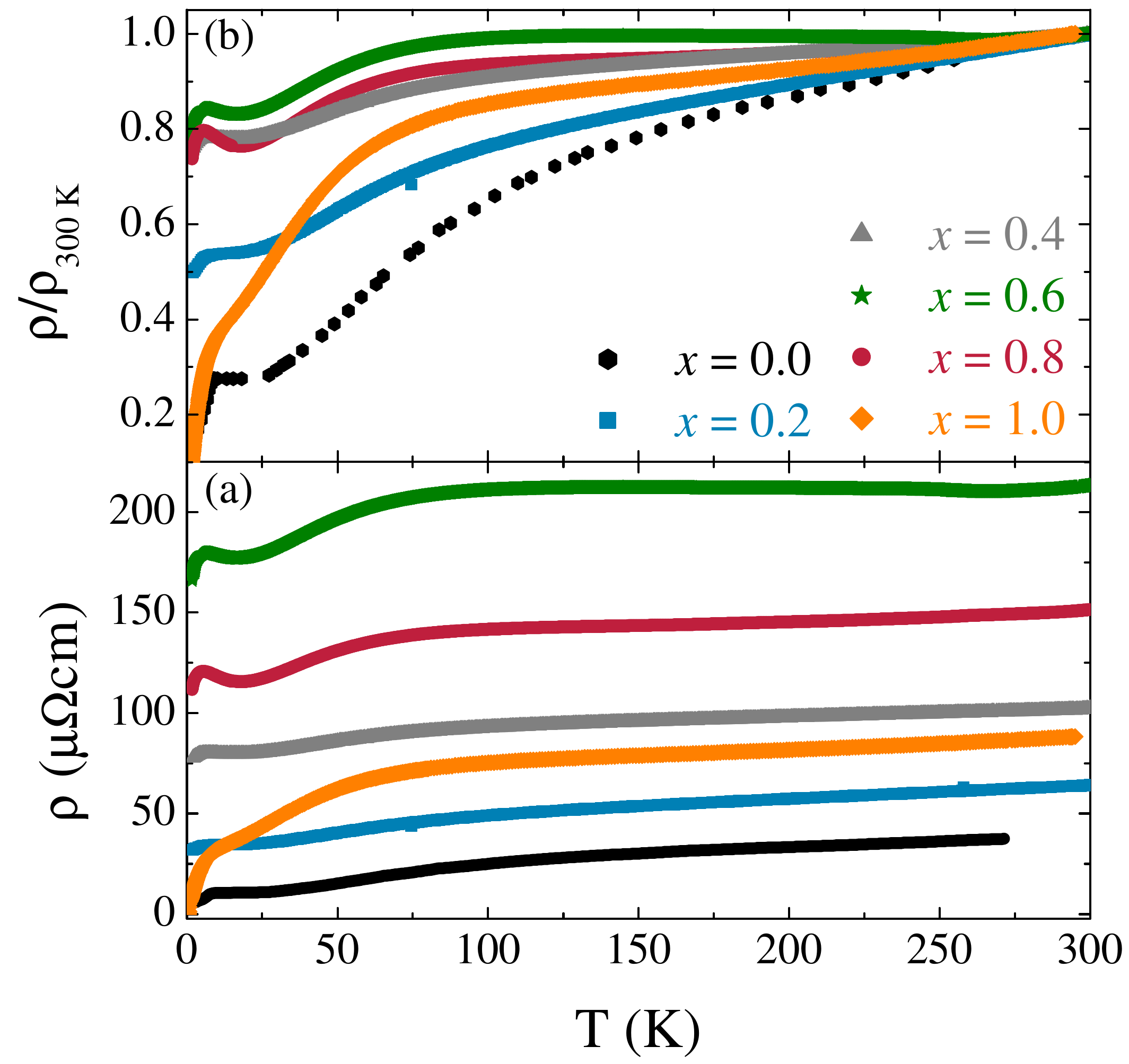}
	\caption{(Color online) Temperature dependence of (a) the resistivity $\rho (T)$ and (b) the normalized resistivity $\rho/\rho_{300 K} (T)$ of CePt$_3$B$_{1-x}$Si$_x$ for $0 \leq x \leq 1$; data for $x = 0$ and 1 taken from Refs.~\cite{sullow,Bauer05}.}
	\label{fig:resistivity}
\end{figure}

In a next step, we have determined the temperature dependence of the electrical resistivity of CePt$_3$B$_{1-x}$Si$_x$ for $0.2\leq x\leq0.8$. In Fig.~\ref{fig:resistivity} the resistivity $\rho(T)$ and the normalized resistivity $\rho/\rho_{300 K}(T)$ are depicted, together with the resistivity data for CePt$_3$B and CePt$_3$Si taken from the Refs.~\cite{sullow,Bauer05}. 

Overall, the absolute values of the resistivities $\rho (T)$ increase with silicon amount up to $x = 0.6$. This behavior reflects the enhanced level of disorder from chemical alloying. Correspondingly, for larger $x$ the disorder level and absolute values $\rho (T)$ decrease.

Further, $\rho (T)$ of the alloying series exhibits a shallow resistive minimum for the intermediate silicon compositions $x = 0.4, 0.6$ and $0.8$ at temperatures of $\sim$ 15 to 20~K. Likely, this behavior results from Kondo scattering at low temperatures.\cite{Kondo} Fits to the temperature range with the resistive upturn using a dependence $\rho - \rho_0 \propto ln(T)$ yield rough estimates for the Kondo temperatures and are included in Tab. \ref{tab:Temp2}. The order of magnitude of the Kondo temperatures is in agreement with the values given for CePt$_3$B and CePt$_3$Si  (see Refs.~\cite{lackner,Bauer05}), although with the small temperature and resistive range fitted no firm conclusions about the $x$ dependence of $T_K$ can be drawn. 

At low temperatures, the drops in $\rho (T)$ denote the transitions into antiferromagnetically ordered states. From the data the antiferromagnetic transition temperatures $T_N$ are determined and included in Tab.~\ref{tab:Temp2}. Consistent with the susceptibility, it is seen that $T_N$ is suppressed with increasing silicon concentration. While the transition of CePt$_3$B occurs as a rather sharp kink, the transitions in alloyed CePt$_3$B$_{1-x}$Si$_x$, $x \geq 0.2$, become broader due to chemical disorder. The transition into the weakly ferromagnetic state at $T_C$ is not observable in the resistivity, in agreement with the conclusions from the Refs.~\cite{sullow,lackner}. Such behavior might be accounted for if at $T_C$ the transition is from one ordered magnetic state into another ordered one, with no change of the size of the magnetic unit cell, and correspondingly no significant changes to the band structure or scattering cross sections.

In an intermediate temperature range, in the (normalized) resistivity a shoulder is observed, denoting scattering from crystalline electric field split levels. The position of the resistive shoulder can be estimated by determining the maximum of the second temperature derivative $d^2 \rho /dT^2$, $T_{CEF}$, and which is a measure for the splitting of the low lying crystal field levels. The compositional dependence of this characteristic temperature $T_{CEF}$ is summarized in Tab. \ref{tab:Temp2}, which decreases with increasing silicon composition. Previously, such crystal field effects have been attributed to a doublet-doublet splitting of the Ce$^{3+}$ ground state (see Refs.~\cite{sullow,lackner}), with a level splitting of the order of magnitude of about 100~K for the different samples. The decrease of the splitting in CePt$_3$B$_{1-x}$Si$_x$ by about 20\% with increasing $x$ is consistent with the increase of the lattice parameter, weakening the electric field strength on the rare earth site.

\begin{figure}[!ht]
	\centering
		\includegraphics[width=1\linewidth]{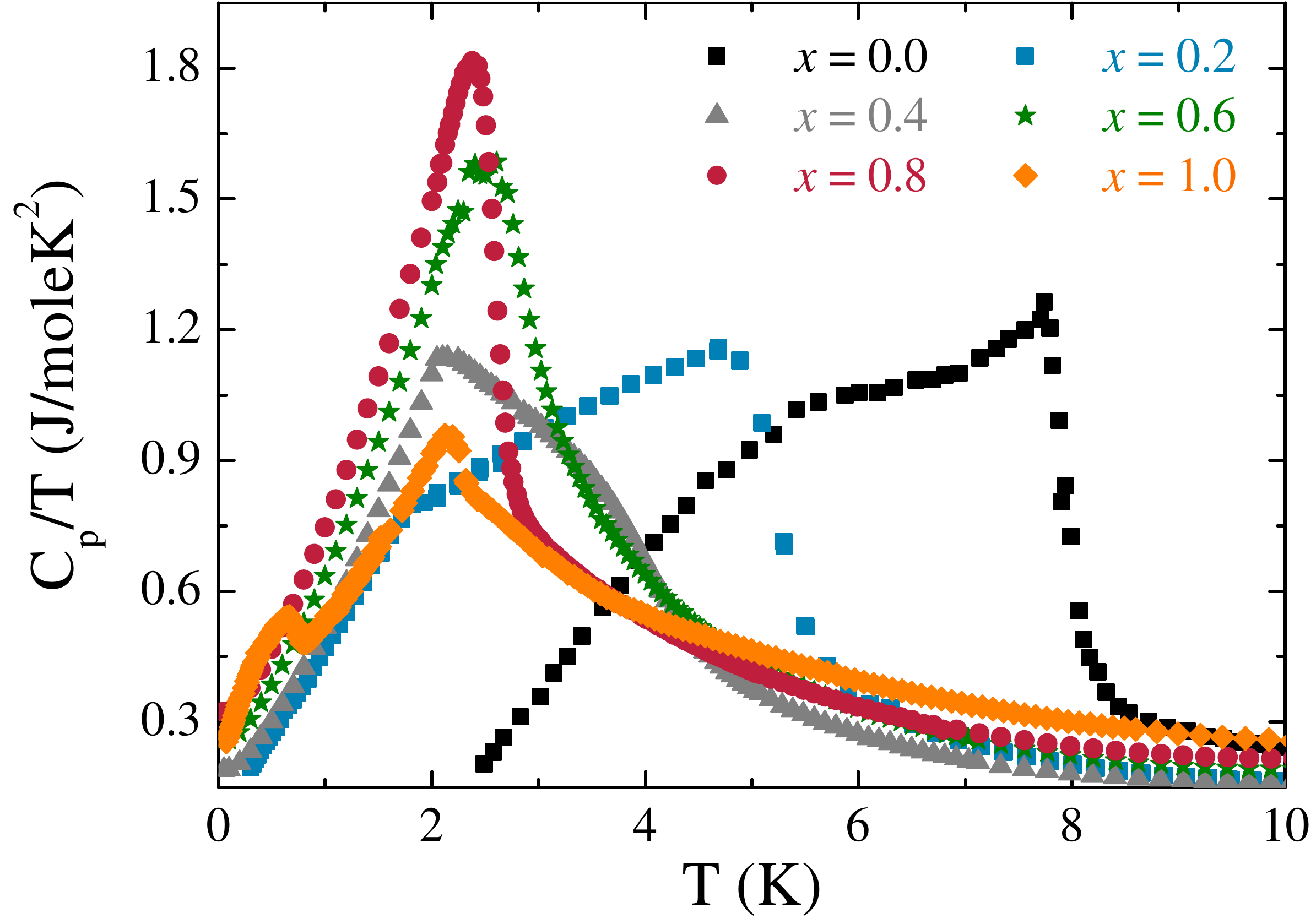}
	\caption{(Color online) Temperature dependence of the specific heat $C_p/T$ versus $T$ of CePt$_3$B$_{1-x}$Si$_x$ for $0 \leq x \leq~1$; data for $x=0$ and $1$ are taken from the Refs.~\cite{sullow,Bauer05}.}
	\label{fig:spezheat}
\end{figure}
\begin{figure}[!ht]
	\centering
		\includegraphics[width=1\linewidth]{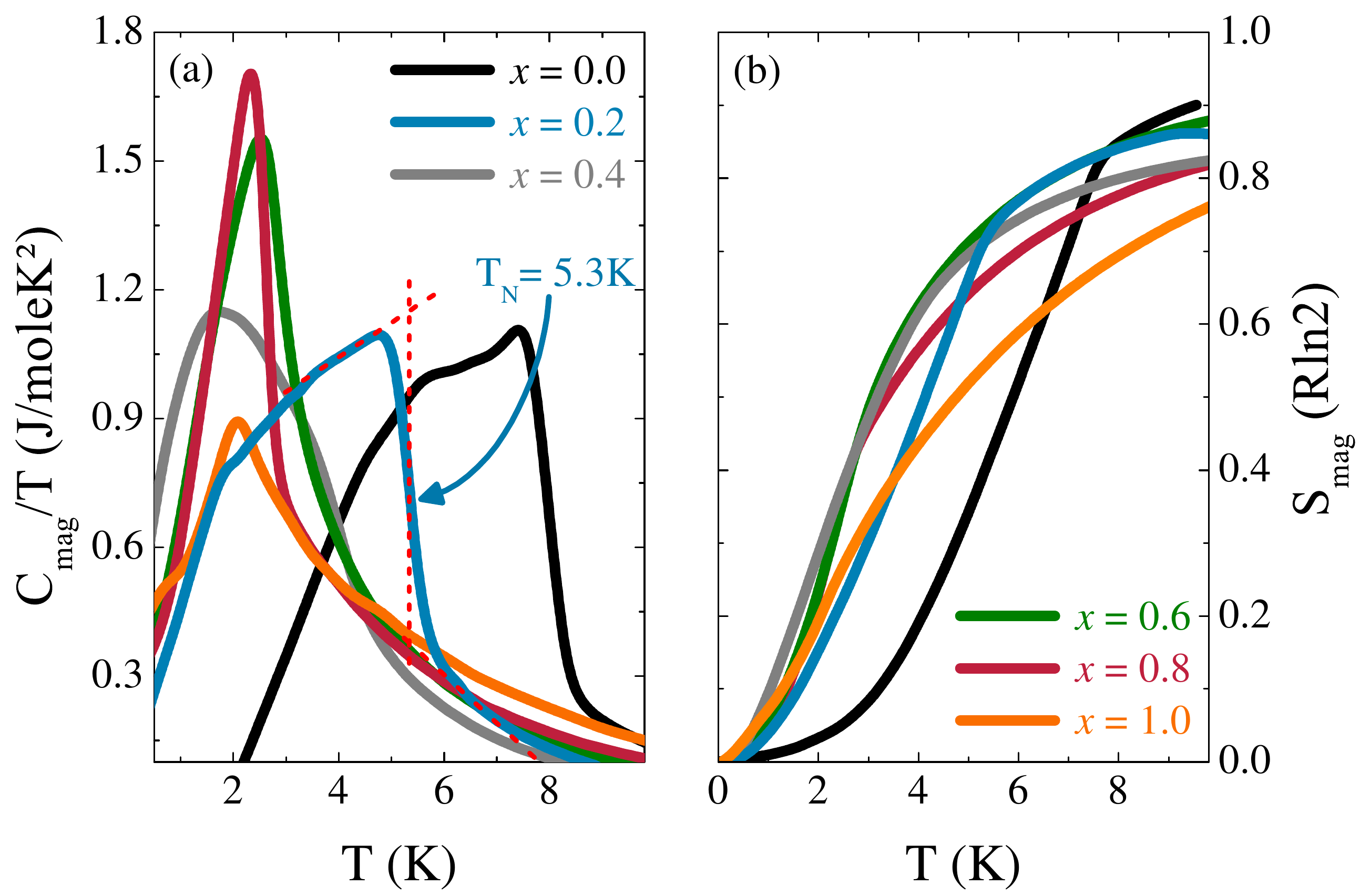}
	\caption{(Color online) (a) Temperature dependent magnetic contribution to the specific heat, $C_{mag}$, plotted as $C_{mag}/T$ versus $T$ and (b) temperature dependence of the magnetic entropy $S_{mag}$ of CePt$_3$B$_{1-x}$Si$_x$ for $0 \leq x \leq~1$. The data for $x=0.0$ and $1.0$ are taken from the Refs.~\cite{sullow,Bauer05}, the line construction illustrates the entropy balance procedure to determine the transition temperature. }
	\label{fig:entropie}
\end{figure}

Finally, in Fig.~\ref{fig:spezheat} we present the specific heat $C_p/T$ versus $T$ at low temperatures. Again, for CePt$_3$B two magnetic phase transitions can be observed as a peak in $C_p/T$ for the antiferromagnetic transition at $T_N$, and as a shoulder for the weakly ferromagnetic transition at $T_C$. On a qualitative level, it is apparent that the main peak is moving to lower temperatures with increasing Si composition.
 To quantify matters and to determine the transition temperatures $T_N$ an entropy balance model is used. For this, in a plot of the magnetic specific heat divided by $T$ ($C_{mag} / T$) versus temperature $T$ the transition temperatures are determined by a linear line construction with equal areas, as it is indicated in Fig. \ref{fig:entropie}(a) for the data of the sample $x = 0.2$. The transition temperatures $T_N$ as determined with this approach are summarized in Tab.~\ref{tab:Temp2}. 
Similarly, the position of the specific heat shoulder at $T_C$ is determined, and which shifts to lower temperatures with increasing silicon amount \cite{shape}. Ultimately, the weakly ferromagnetic phase is not perceivable anymore for $x>0.6$. Correspondingly, for CePt$_3$Si only the antiferromagnetic peak is identified, implying that ferromagnetism has disappeared for large values of $x$.  

In a next step, from the experimental specific heat data the magnetic contribution is derived by subtracting the lattice contribution of isostructural, non-magnetic LaPt$_3$B$_{1-x}$Si$_x$. For this, we use the experimentally determined specific of LaPt$_3$B and LaPt$_3$Si (see Refs.~\cite{Bauer04,lackner}) and calculate the corresponding lattice contributions for alloyed LaPt$_3$B$_{1-x}$Si$_x$ from an interpolation of the specific heat of the two end points. This way, we obtain the magnetic specific heat contribution $C_{mag}$ plotted in Fig.~\ref{fig:entropie}(a) as $C_{mag} / T$ vs. temperature $T$. From this plot, clearly the change of the shape of the specific heat curve is seen from a double-peak like structure for $x \leq 0.6$ to a single antiferromagnetic peak for $x \geq 0.8$.

Previously, for CePt$_3$B the magnetic specific heat contribution at low temperatures (in the antiferromagnetic state) was described in terms of spin wave excitations following the model of Continentino {\itshape et al.}\cite{Continentino},
\[
	C_{mag} = \frac{\delta \Delta^{7/2} \sqrt{T}}{\exp\left(\frac{\Delta}{T}\right)}\left[1+\frac{39}{20}\left(\frac{T}{\Delta}\right)+\frac{51}{32}\left(\frac{T}{\Delta}\right)^{2}\right],
\]
with $\delta \propto 1/D^3$. Here, $D$ is the spin wave velocity, while $\Delta$ represents the value of the antiferromagnetic spin wave dispersion gap. Moreover, the electronic contribution to the specific heat $\propto \gamma T$ is also taken into account, with the Sommerfeld coefficient $\gamma$. Fits of $C_{mag}/T$ deliver a $x$ dependence of the Sommerfeld coefficient $\gamma$ summarized in Tab.~\ref{tab:Temp2}.\cite{footnote2} Evidently, with increasing $x$ the Sommerfeld coefficient and the effective electron mass $m^*$ increase as well, thus reflecting a transition from a local moment antiferromagnet (CePt$_3$B) to a heavy fermion system (CePt$_3$Si) upon alloying.

With the magnetic specific heat extrapolated this way to $T = 0$\,K we can calculate the magnetic entropy $S_{mag}$ (Fig.~\ref{fig:entropie}(b)). Consistent with the enhancement of the Sommerfeld coefficient with $x$, we find a suppression of the entropy recovered at $T_N$, with $S_{mag}$ at $T = T_N$, measured in units $R \cdot \ln{2}$, summarized in Tab.~\ref{tab:Temp2}. Commonly, the magnetic entropy $S_{mag}$ is considered to scale with the size of the ordered magnetic moment. Given that in CePt$_3$Si there is an ordered magnetic moment of $\mu_{ord} = 0.16 \mu_B$, and taking into account a reduction of the magnetic entropy $S_{mag} (T = T_N)$ by a factor of four from CePt$_3$B$_{1-x}$Si$_x$, $x = 1$ to $x = 0$, it would suggest that for CePt$_3$B the ordered moment should be of the order of 0.6~$\mu_B$. This finding is consistent with our observations on the bulk properties of CePt$_3$B$_{1-x}$Si$_x$ as well as the results of muon spin rotation experiments from Ref.~\cite{rauch}.

\section{Discussion and Conclusion}

\begin{figure}[!ht]
	\centering
		\includegraphics[width=1\linewidth]{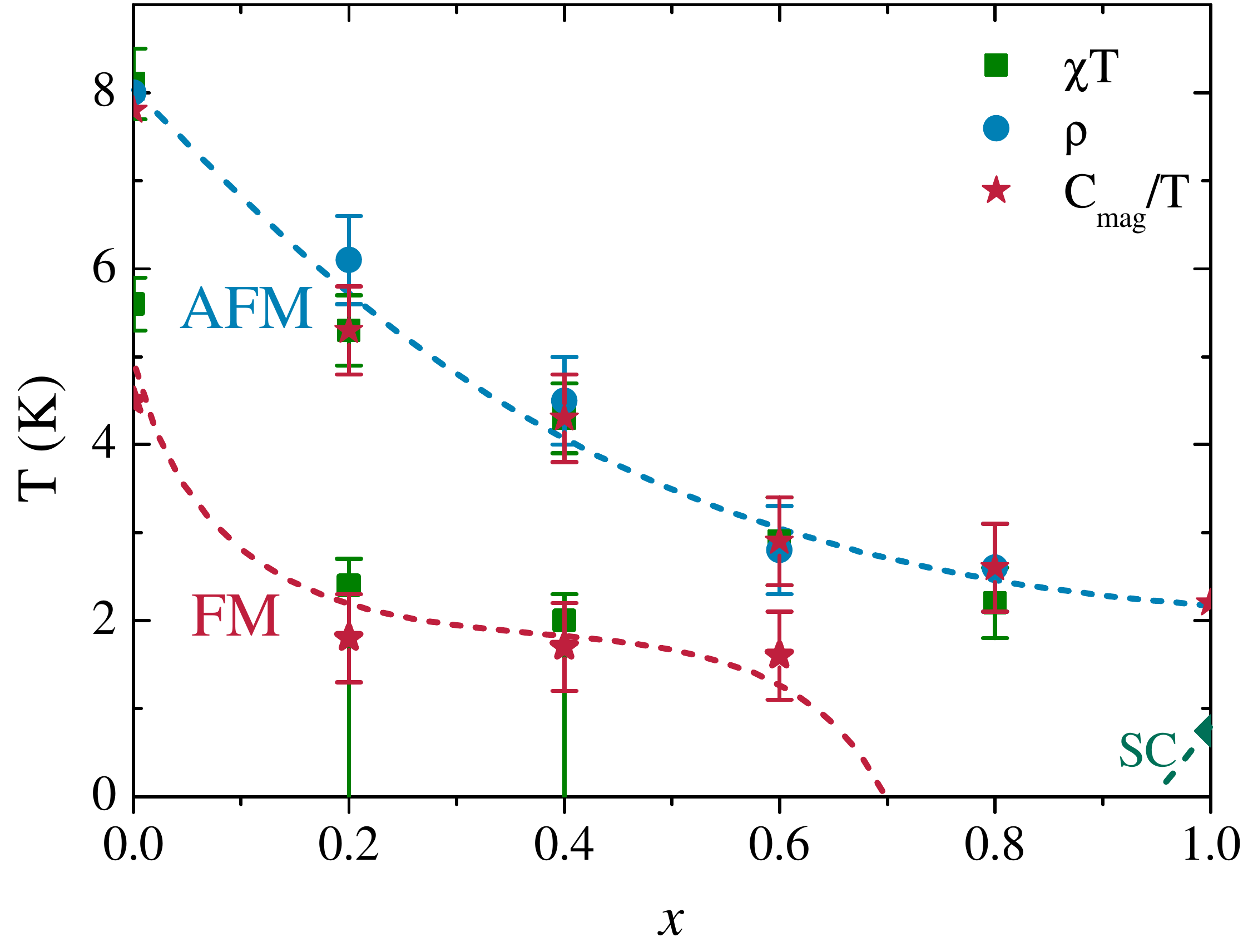}
	\caption{(Color online) Magnetic phase diagram of the antiferromagnetic $T_N$, ferromagnetic $T_C$ and superconducting $T_c$ transition temperatures in dependency of the concentration of silicon, $x$, for CePt$_3$B$_{1-x}$Si$_x$, $0 \leq x \leq~1$; data for $x = 0$ and 1 taken from Refs.~\cite{sullow,Bauer05}.}
	\label{fig:phasedia}
\end{figure}

With the values of the antiferromagnetic and ferromagnetic transition obtained from various experimental techniques and summarized in Tab.~\ref{tab:Temp2} we construct the magnetic phase diagram depicted in  Fig.~\ref{fig:phasedia}. While there is some variation of the absolute values of $T_N$ and $T_C$ derived from different techniques, overall we find a continuous transformation of the antiferromagnetic phase in CePt$_3$B into that of CePt$_3$Si, with a smooth suppression of ordering temperatures. In contrast, the weakly ferromagnetic phase in CePt$_3$B is completely suppressed at a critical value of $x \sim 0.7$. Eventually, superconductivity appears close to stoichiometric CePt$_3$Si, although from our data we cannot accurately determine the critical concentration $x_c$ of the appearance of superconductivity.

The magnetic phase diagram has to be discussed in context with the evolution of the other physical parameters of the alloying series CePt$_3$B$_{1-x}$Si$_x$. The suppression of magnetic order is accompanied by a significant enhancement of electronic correlations, as evidenced by the increasing electronic specific heat coefficient $\gamma$. Qualitatively, this behavior can be discussed within the concept of the Doniach phase diagram, which considers the competition of long range magnetic order from an RKKY-like exchange and the Kondo effect.\cite{doniach} Here, the basic idea is that there is a difference in the dependence on the magnetic coupling strength $J$ of the characteristic energy scales of the RKKY-exchange, $k_B T_{RKKY} \propto J^2 N(E_F)$ ($N(E_F) =$ density of states at the Fermi energy), and the Kondo energy, $k_B T_{Kondo} \propto (1/N(E_F)) \exp \left[-1/(J N(E_F)\right]$. For comparatively small $J$ values the RKKY-exchange dominates, causing magnetically ordered local moment states to occur. In contrast, for large $J$ values the Kondo screening will prevail, and long-range order will be suppressed. 

The resistivity and the $x$ dependence of the Curie-Weiss temperature indicate that for both CePt$_3$B and CePt$_3$Si the Kondo energy scale is of the order 10\,K, and increases with $x$ by about a factor of two. Further, the Curie-Weiss temperature $\Theta_{CW}$ indicates that the basic magnetic energy scale $T_{RKKY}$ is slightly larger than $T_K$ (25 to 45\,K) and which increases with $x$, too. In result, with replacing B by Si in CePt$_3$B$_{1-x}$Si$_x$ a Doniach-like phase diagram is traversed, starting with the local moment magnet CePt$_3$B. For increasing Si content $x$ the Kondo effect is enhanced and tends to win over magnetic order, with the end point of the heavy fermion antiferromagnetic superconductor CePt$_3$Si.

Commonly, such Doniach-like phase diagrams are observed in pressure experiments or in isoelectronic chemical pressure studies. In our case, the situation is somewhat more complicated. Chemically a negative pressure is exerted in the alloying series CePt$_3$B$_{1-x}$Si$_x$ with a large increase of the lattice parameters with $x$. The negative pressure effect is also corroborated by the observation of a decreasing crystal field splitting with $x$. This negative pressure, however, appears to be counteracted by the increasing electron count while replacing B by Si. Adding one conduction electron might lead to a small shift of the Fermi energy, thus affect the density of states at the Fermi level, $N(E_F)$. However, as first approximation the replacement of B by Si is often considered to not significantly change band structure properties.\cite{sandratskii,divis,yaresko} These elements only produce flat and broad bands at the Fermi level, and which should not be of great relevance to the magnetic properties of the materials. In consequence, as result of our experimental study, CePt$_3$B appears to represent a low pressure variant of CePt$_3$Si. 

In fact, pressure experiments carried out so far on CePt$_3$B are broadly consistent with this statement.\cite{lackner} Here, for pressure up to 1.85\,GPa only a slight increase of $T_N$ by about 10\% is observed, without qualitative changes to the character of the magnetic ground state. Within the concept of the Doniach model, this observation would reflect that CePt$_3$B is still deep in the local moment region of the Doniach phase diagram, and that a much larger pressure would be required to drive the system into the range of strong electronic correlations and close to a magnetic instability. Consequently, it would be very interesting to see if the properties of CePt$_3$B under very high pressure resemble those of CePt$_3$Si, and in particular if the system becomes superconducting.

There are a couple of further points to be considered in context with the phase diagram of CePt$_3$B$_{1-x}$Si$_x$. First, there is the non-observation of superconductivity for CePt$_3$B$_{1-x}$Si$_x$, $x = 0.8$, down to 0.3\,K. Strictly speaking, to ultimately settle this experimental case, it would be necessary to carry out experiments to lower temperatures ($^3$He/$^4$He) as well as for samples CePt$_3$B$_{1-x}$Si$_x$ with values $x$ closer to 1. Conceptually, however, the rapid suppression of superconductivity by doping with non-magnetic elements is in line with the ideas and observations about unconventional superconductivity in general (see Refs.~\cite{millis,dalichaouch,sullow2,mackenzie}) and in CePt$_3$Si specifically.\cite{nicklas,grytsiv} 

Secondly, there is the issue about the relationship between the antiferromagnetic states in CePt$_3$B and CePt$_3$Si. Based on our observations, we conclude that the simple antiferromagnetic structure of CePt$_3$Si, with ${\bf q} = (0, 0, 0.5)$, evolves out of the antiferromagnetic phase of CePt$_3$B. For the latter compound, however, it has been demonstrated that the AFM structure does not consist of a ${\bf q} = (0, 0, 0.5)$ ordering,\cite{rauch} but instead possesses a different (possibly rather complex) magnetic structure. The question arises, if there is a continuous transformation of the magnetic structure in CePt$_3$B into that of CePt$_3$Si, or   if it is instantaneous, and how this does affect the magnetic fluctuation spectrum. This, in turn, might be of relevance to the issue of the coupling mechanism of unconventional superconductivity in CePt$_3$Si. Similarly, the suppression of weak ferromagnetism with $x$ might be associated to modifications of the magnetic fluctuation spectrum relevant to superconductivity.

\section*{ACKNOWLEDGMENTS}

Parts of this work were supported by the Austrian FWF (P22295), the Laboratory for Magnetic Measurements (LaMMB) at HZB and the Japanese Society for the Promotion of Science. Work at Stony Brook University is supported by the National Science Foundation under grant DMR-0907457.


\begin{references}
\bibitem{footnote1} The correction $\chi_0$ to the Curie-Weiss law is minor, with $\chi_0$ up to max. $5$~\% of $\chi_{300K}$(T).
\bibitem{shape} Qualitatively, the sample $x = 0.4$ has a different appearance than $x = 0.2$, with the maximum in $C_p/T$ now occurring at $T_C$ rather than $T_N$. Apparently, this change of shape reflects that for this sample entropy is shifted from the antiferromagnetic phase to the ferromagnetic one, if compared to the sample $x = 0.2$. Since we have no microscopic information about type of the ordered states nor size of the magnetic moments, we cannot draw any firm conclusions about a relationship between shape of the specific heat and the magnetically ordered state.
\bibitem{footnote2} Because of the rather complex fit of the low temperature specific heat there is an estimated uncertainty in the determination of $\gamma$ of about 5 \% resulting from parameter interdependency.
\end{references}
\end{document}